\documentclass[cup6b]{cupbook}
\usepackage{graphicx}
\usepackage{natbib}
\title[Rome, Italy, 27--30 April 2009]
      {The coming of age of X-ray polarimetry}
\author{}
\date{}
\begin{document}
\pagenumbering{arabic}


\author[N. Barri\`ere et al.]
{N. Barri\`ere, L. Natalucci and P. Ubertini  \and (INAF - IASF Roma, via Fosso del Cavaliere 100, 00133 Roma, Italy)}

\chapter{Hard X / soft gamma ray polarimetry using a Laue lens}

%
\abstract{Hard X / soft gamma-ray polarimetric analysis can be performed efficiently by the study of Compton scattering anisotropy in a detector composed of fine pixels. But in the energy range above 100 keV where sourcesÕ flux are extremely weak and instrumental background very strong, such delicate measurement is actually very difficult to perform. 
Laue lens is an emerging technology based on diffraction in crystals allowing the concentration of soft gamma rays. This kind of optics can be applied to realize an efficient high-sensitivity and high-angular resolution telescope, at the cost of a field of view reduced to a few arcmin though.
A 20 m focal length telescope concept focusing in the 100 keV - 600 keV energy range is taken as example here to show that recent progresses in the domain of high-reflectivity crystals can lead to very appealing performance. The Laue lens being fully transparent to polarization, this kind of telescope would be well suited to perform polarimetric studies since the ideal focal plan is a stack of finely pixelated planar detectors - in order to reconstruct the point spread function - which is also ideal to perform Compton tracking of events.}

%
\section{Introduction}
A Laue lens concentrates gamma-rays using Bragg diffraction in the volume of a large number of crystals arranged in concentric rings and accurately orientated in order to diffract radiation coming from infinity towards a common focal point (e.g. \cite{lund.92, halloin.05}). This principle is applicable from $\sim$ 100 keV up to 1.5 MeV, but with a unique lens it is difficult to cover efficiently a continuous energy band of more than about half an order of magnitude wide. 
Thanks to the decoupling between collecting area and sensitive area, a Laue lens produces a dramatic increase of the signal to background ratio. This is the key to achieve the so awaited sensitivity leap of one - two orders of magnitude with respect to past and current instruments (IBIS and SPI onboard INTEGRAL, Comptel onboard CGRO,  BAT onboard SWIFT).
 
Despite a Laue lens is a concentrator (i.e. not a direct imaging system) it allows the reconstruction of images with an angular resolution of $\sim$ 1 arcmin in a field of view of $\sim$ 10 arcmin.
Moreover, combined with a focal plane capable of tracking the various interaction of an event, the telescope becomes sensitive to the polarization:  the lens being fully transparent to polarization \cite{barriere.09, curado-da-silva.08} it lets the possibility to analyze it in the focal plane. Using the non-uniformity of the azimuthal angle of Compton diffusion, it becomes possible to perform a statistical study of the signal in order to measure its polarization angle and fraction.

A telescope featuring a Laue lens would be perfectly adapted to further our understanding of high-energy processes occurring in a variety of violent events. High sensitivity investigations of point sources such as compact objects, pulsars and active galactic nuclei should bring important insights into the still poorly understood emission mechanisms. For this purpose the polarization detection capabilities inherent to a Laue lens telescope in conjunction with its high angular resolution would be extremely helpful, for instance to distinguish between various models of pulsars magnetosphere. Also the link between jet ejection and accretion in black hole and neutron star systems could be clarified by observation of the spectral properties and polarization of the transition state emission.

%
\section{Proposed telescope}

The main driver of the studied concept is to perform high-sensitivity hard X-ray pointed observations in the energy range 100 keV- 600 keV. The Laue lens, which have a focal distance of 20 m, can be maintained by extensible booms allowing a medium sized mission to carry the telescope.

%
\subsection{Laue lens}
\begin{figure}[t]
\begin{center}
\includegraphics[width=0.59\textwidth]{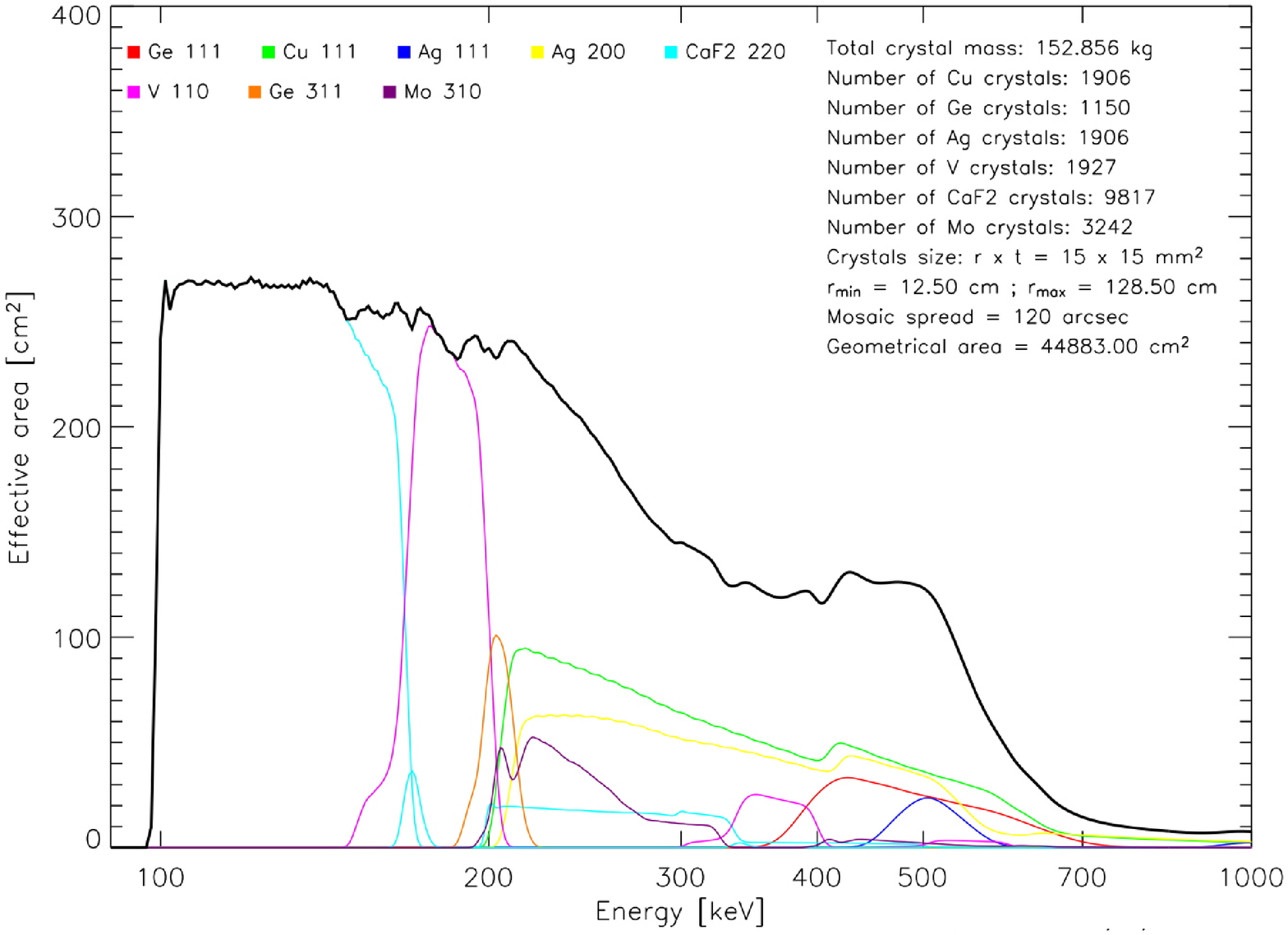}
\includegraphics[width=0.40\textwidth]{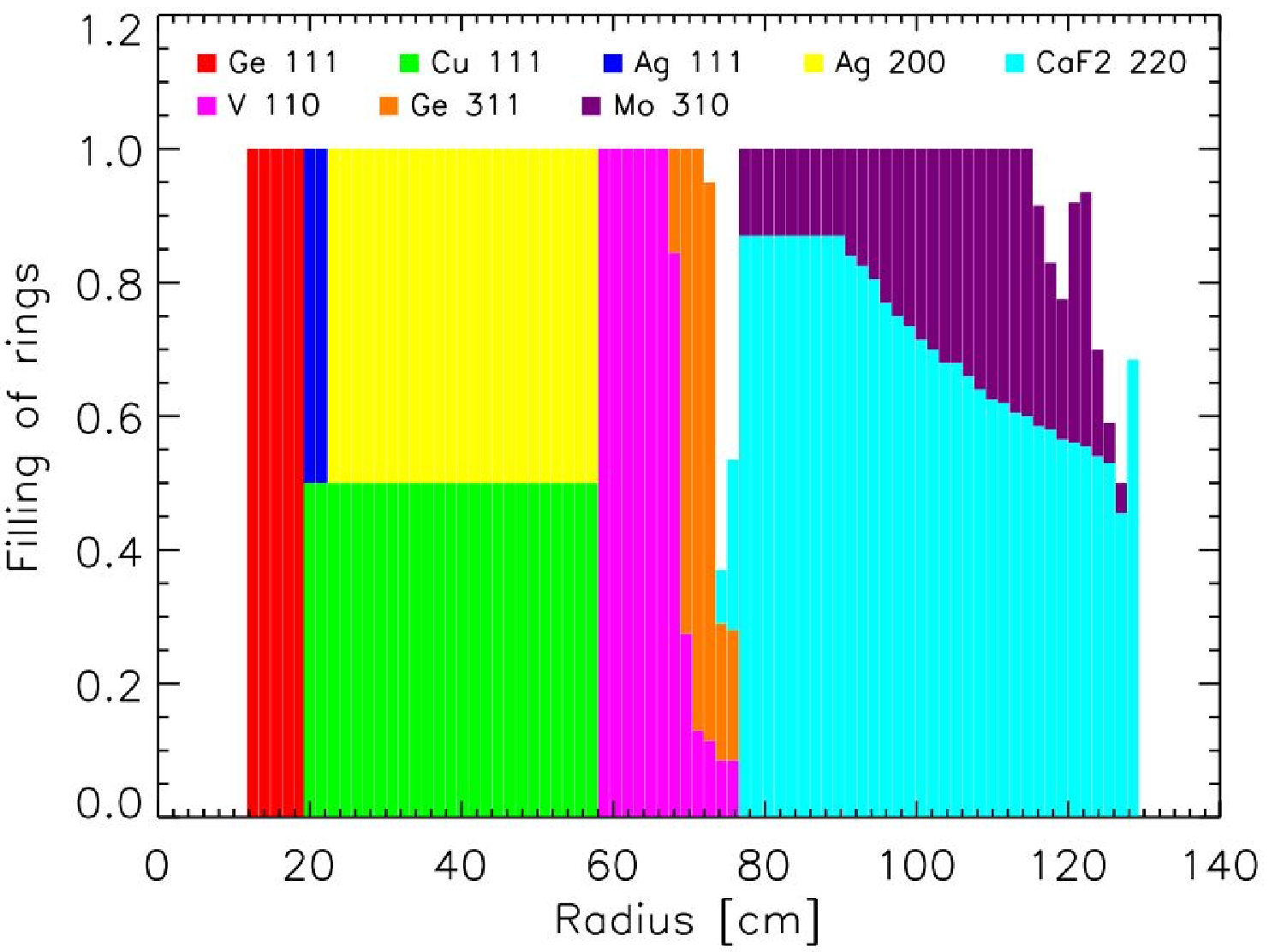}
\caption{\textit{Left.} Effective area of the lens. \textit{Right.} Reparation of the various crystals and reflections used to reach the effective area.}
\label{fig:Lauelens}
\end{center}
\end{figure}

The Laue lens designed for this study is composed of 20000 crystal tiles of 15 mm x 15 mm in a complex assembly of CaF2, Cu, Ge, Ag, V and Mo. The required mosaicity for the crystal is 2 arcmin, which is not a too tight constraint for their production (growth and cut) and mounting. Figure \ref{fig:Lauelens} shows the effective area of the lens and the repartition of the various crystals over the lens radii. Crystals development for Laue lenses is currently addressed in various institutes in Europe, and gives very positive results \cite{ferrari.08, barriere.09b}. Crystals total mass equals 150 kg, and based on MAX \cite{duchon.05} and GRI \cite{knodlseder.09} mission studies, its structure can be estimated to $\sim$ 70 kg/m$^2$, which make a total weight of 460 kg for the lens.

%
\subsection{Focal plane instrument}

The advantages of having a Compton camera at the focus of the lens are multiple. The inherent fine pixelisation allows the reconstruction of the point spread function (PSF) via the localization of the first interactions, which is fundamental to take fully benefit of the lens. Since the signal is confined in a defined area, simultaneous monitoring of the background is possible using the non-lighted detector areas allowing efficient evaluation of its time variability.
Background rejection can be further enhanced if Compton kinematics is applied to the reconstruction of events. At the cost of a non-negligible fraction of the signal, it is possible to constrain the original direction of the events within an annulus, and in many cases, to establish wether the photon energy was fully recorded or not \cite{zoglauer.07}. This makes a powerful tool to reject efficiently background by discriminating photons whose 'event circle' does not intercept the lens direction. 

Performing Compton  a statistical number of event reconstructions directly allows the measurement of the signal polarization angle and fraction. However it is necessary that the first layer is made out of silicon (Si) to apply Compton kinematics down up 100 keV. At 100 keV Compton interaction is by a factor of $\sim$ 10 the most probable interaction in Si, but the drawback is it has a low interaction cross section:  Mean free path at 100 keV in Si is already 2.5 cm. Consequently Si layers have to be completed by additional layers of a much more absorbent material of which CdTe (or CdZnTe) is of prime choice \cite{takahashi.06, ubertini.03}.

A focal plane of 8 cm x 8 cm gives a field of view of 10 arcmin FWHM (limited by the spread of the signal over the focal plane that induces a sensitivity loss).  For point sources, the reconstruction of the PSF allows the localization to less than 1 arcmin within the field of view. In case of an extended source, combined observations in a dithering pattern still permit to take advantages of the imaging capabilities.

%
\section{Conclusion}
According to preliminary studies, continuum sensitivity better than 2x10$^{-7}$ ph/s/cm$^2$/keV (3$\sigma$, 10$^5$ s, $\Delta$E/E = E/2, point source on-axis) is reachable using the Laue lens described in this paper. This sensitivity could allow polarimetric studies of bright sources with an unprecedented angular resolution (especially very interesting for the Crab nebula).
An appealing option to complete the Laue lens telescope would be to perform all-sky monitoring covering half the sky continuously over the entire mission lifetime. Thanks to a dedicated design it would be conceivable that the instrument at the focus of the lens behaves simultaneously as focal plane and large field of view Compton camera.   
%
%
\section*{Acknowledgement}
NB and LN are grateful to ASI for the support of Laue lens studies through grant I/088/06/0.


\end{document}